\documentclass[10pt]{iopart}
\usepackage{array}
\usepackage{graphicx}
\usepackage{svg}
\usepackage{amssymb}
\expandafter\let\csname equation*\endcsname\relax
\expandafter\let\csname endequation*\endcsname\relax

\usepackage{url} 

\usepackage{soul, xcolor}
\setstcolor{red}
\usepackage[thinc]{esdiff}
\usepackage{amsmath}
\usepackage{mathtools}
\usepackage{array} 
\usepackage{xr} 
\usepackage{float} 
\usepackage{array} 

\usepackage{cite}

\begin{document}
\pagestyle{plain} 

\title[Automated tuning and characterization of single-electron and single-hole transistor charge sensors]{Automated tuning and characterization of single-electron and single-hole transistor charge sensors}

\author{B. Van Osch$^{1, 2}$,  A. Paurevic$^{1, 2}$,  A. Sakr$^3$,  T. Joshi$^{1,5}$, D. van der Bovenkamp$^3$, Q. T. Nicolau$^3$, F. A. Zwanenburg$^3$, J. Baugh$^{1, 2, 4, 6}$\footnote{Author to whom any correspondence should be addressed.}}

\address{$^{\text{1)}}$ Institute for Quantum Computing, University of Waterloo, Waterloo N2L 3G1,
Canada}
\address{$^{\text{2)}}$ Department of Physics, University of Waterloo, Waterloo N2L 3G1, Canada}
\address{$^{\text{3)}}$ MESA+ Institute for Nanotechnology,
University of Twente, PO Box 217, 7500 AE Enschede, The Netherlands}
\address{$^{\text{4)}}$ Department of Chemistry, University of Waterloo, Waterloo N2L 3G1, Canada}
\address{$^{\text{5)}}$ Department of Electrical and Computer Engineering, University of Waterloo, Waterloo N2L 3G1, Canada}
\address{$^{\text{6)}}$ Waterloo Institute for Nanotechnology, University of Waterloo, Waterloo N2L 3G1, Canada}
\ead{baugh@uwaterloo.ca}

\vspace{12pt}

\begin{abstract}
We present an automated protocol for tuning single-electron transistors (SETs) and single-hole transistors (SHTs) to operate as high-sensitivity DC charge sensors. The protocol initializes a previously unmeasured device after cooldown, identifies a working point in barrier-gate space, and selects and ranks charge-sensing operating points. It further automates the acquisition and analysis of Coulomb diamonds to extract sensor-relevant parameters, including lever arm, charging energy, gate and source/drain capacitances, and estimated dot radius. We demonstrate the protocol on accumulation-mode silicon MOS SET and SHT devices operated at 1.5~K and \(\approx 50\)~mK, respectively, establishing ambipolar applicability across a wide temperature range. Operation at 1.5~K indicates that charge sensing in compact MOS devices is feasible in the 1--2~K regime, supporting higher-temperature readout relevant to scalable spin-qubit architectures. Compared to manual tuning, automation reduces operator overhead and provides consistent device characterization, with clear pathways for further speedups and improved robustness via faster electronics and feedback-based stabilization.
\end{abstract}

\vspace{2pc}

\maketitle

\section{Introduction}

Single-electron transistors (SETs) and single-hole transistors (SHTs) have demonstrated significant utility as precision charge sensors with applications in quantum devices, metrology, and fundamental physics. This includes readout of semiconductor spin qubits based on spin-to-charge conversion, for which radio-frequency SETs (RF-SETs) are particularly effective for fast, high-fidelity charge sensing \cite{RF-SET-Angus, RFSET_qubits}. Other examples include realization of quantum current standards \cite{metrology}, and detection of fractional charge excitations in the quantum Hall effect \cite{fractional}. In order to act as a charge sensor, a SET device must be tuned to an operating point where its conductance is highly sensitive to changes in the local electrostatic environment. While sensitivity tuning can be done heuristically, there have been recent efforts efforts to automate tuning of charge sensors and gate-defined quantum dot devices more broadly, in order to speed up experiments and improve reliability \cite{Colloquium_Zwolak_2023}. 

Different automation protocols have targeted various phases of the tuning process, often with an end goal of tuning quantum dots for operation as spin qubits. Initialization methods establish a suitable range of gate voltages for conducting channel formation, tunnel barrier pinch-off, and forming the desired number of quantum dots in the system \cite{McJunkin:2021ros, Darulov__2020}. Further tuning involves adjusting the number of charges on each dot \cite{Baart_2016, Yon_2024,  Moon_2020}, setting up virtual gating, and fine tuning for qubit manipulation and readout \cite{Schuff2026Autonomous}. 

Beyond electron-based spin qubits, there is growing interest in ambipolar devices which make use of both SET and SHT charge sensors, due to the potential advantages of hole quantum dots as spin qubits. Notably, the large spin-orbit coupling of holes enables efficient spin manipulation using electric dipole spin resonance (EDSR) \cite{hole_qubit,EDSR}. Coherent shuttling has been explored extensively for electrons \cite{shuttling}, motivating hybrid concepts where holes could serve as static nodes while electrons mediate transport. Ambipolar devices allow for characterizing co-located electron and hole quantum dots. As demonstrated by Sousa de Almeida \emph{et. al.} \cite{ambipolar}, closely spaced SET and SHT structures have been used for mutual charge sensing, using one device as a charge sensor for the other.  

In this work, we present an automated protocol \cite{qdot_control} that tunes both electron- and hole-based charge sensing devices. The tuning protocol initializes devices after cooldown using only generic device layout information, and determines several key characteristics of the sensor, such as the estimated dot radius and gate lever arm based on analysis of Coulomb diamonds. We demonstrate tuning of an accumulation mode SET device at a temperature of 1.5~K, where the device is sufficiently compact to allow its charging energy to exceed the thermal energy scale. This shows the effectiveness of the method in the 1-2~K temperature range, where ``hot" spin qubits could be operated in future large-scale quantum computing devices \cite{hotdensecoherent}. The protocol is further applied to an SHT device operated at dilution refrigerator temperatures, demonstrating its ambipolarity and applicability across temperature ranges. Compared to prior automation efforts primarily aimed at qubit-dot formation and charge-state control, our focus is automated initialization of MOS SET/SHT charge sensors from cooldown, plus automated Coulomb-diamond characterization for sensor-relevant calibration and screening. 

\section{Experimental Setup}
\label{sec:device}

\begin{figure}[h]
    \centering
    \includegraphics[width=\linewidth]{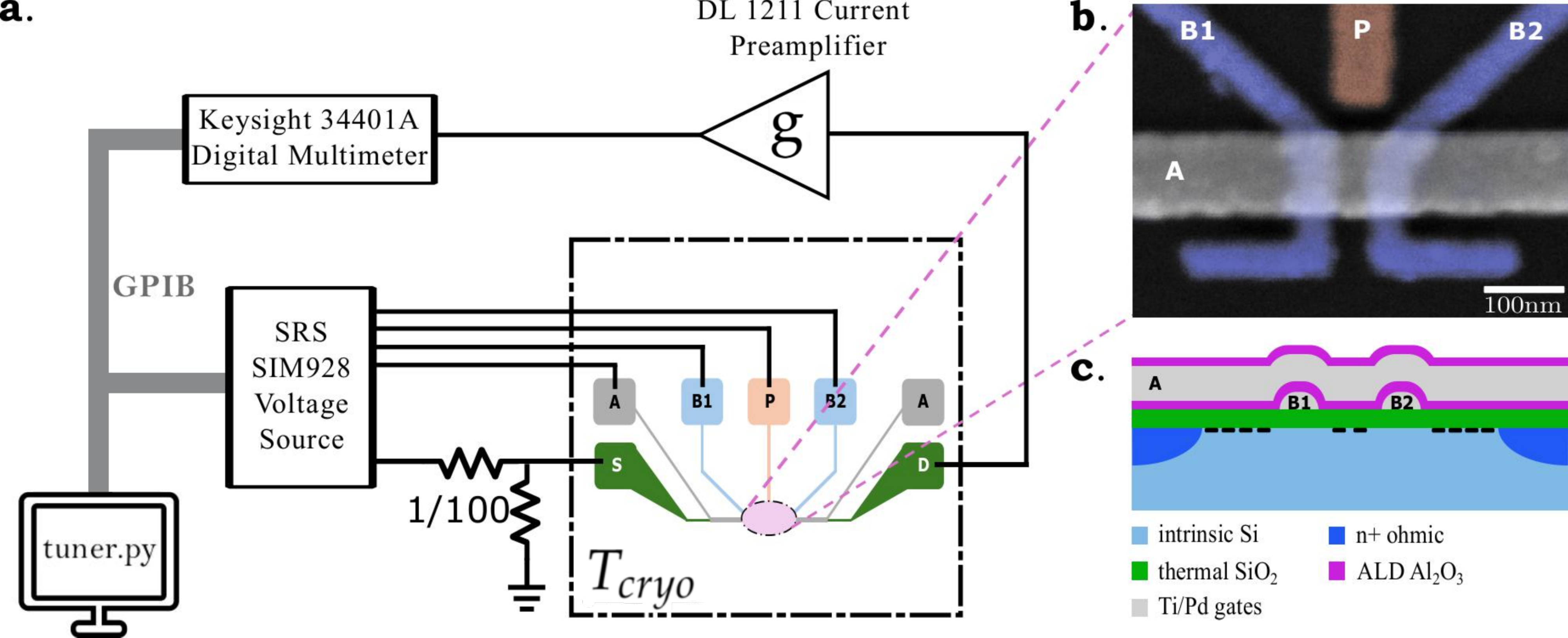}
    \caption{\textbf{(a)} Schematic of the experimental setup. The device is held at $T_{cryo}$, which is either 1.5~K in a pumped $^4$He cryostat or $\approx 50$ mK in a dilution refrigerator. The computer, DC voltage source and digital multimeter communicate via GPIB. A Python script (tuner.py) runs the tuning protocol by controlling the voltages supplied to the device gate, source and drain electrodes using an SRS SIM928 voltage source. The data is then acquired from a Keysight 34401A multimeter. The source-drain current is read by a current-voltage preamplifier fed into the digital multimeter. The preamplifier gain (g) is typically set in the range of $10^7-10^8$ V/A. \textbf{(b)} False-color scanning electron micrograph (top view) and \textbf{(c)} schematic cross-section (side view) of the SET device. The Ti/Pd gate electrodes are deposited by electron-beam physical vapor deposition. Separating the barrier gates B1 and B2 from the top accumulation gate A is a 5 nm film of Al$_2$O$_3$ deposited by atomic layer deposition. The conducting channel forms in intrinsic (undoped) silicon, below 10 nm of thermally grown SiO$_2$. The doped (n$^{+}$) Ohmic regions shown in \textbf{(c)} are only schematic; they are far away from the device region in the actual device. The SHT device has a similar structure and geometry, but with p$^{+}$-doped Ohmic contacts. }
    \label{fig:setup}
\end{figure}

The automated tuning protocol is first demonstrated on a silicon metal-oxide-semiconductor (MOS) SET cooled to 1.5~K in a pumped helium-4 ($^4$He) cryostat. Figure~\ref{fig:setup} shows a schematic of the experimental setup. The SET has two layers of gate electrodes, separated by a dielectric. The accumulation gate A (top layer) induces a 2DEG, as it overlaps with degenerately doped (n$^{++}$) regions formed by shallow ion implantation far from the active device area that act as Ohmic contacts. In the device active area, the induced 2DEG is locally depleted by barrier gates B1 and B2 (bottom layer) to form tunneling barriers. A quantum dot forms between the two barrier gates, with its electrochemical potential tuned by the plunger gate P as well as the barrier gates. 

DC voltages are supplied by a SIM928 Isolated Voltage Source (Stanford Research Systems) assembled on a SIM900 mainframe for the measurements conducted at 1.5 K, and by an IVVI Rack (QuTech) for the $\approx 50$ mK measurements. A source-drain DC voltage of 100 $\mu$V is applied to the device through a voltage divider. The SET/SHT current is measured by a Model 1211 Current Preamplifier (DL Instruments). The output voltage of the current preamplifier is read by a digital multimeter, which sends the acquired data to the computer.

\section{Tuning Protocol}
\begin{figure}[ht!]
    \centering
    \includegraphics[width=\linewidth]{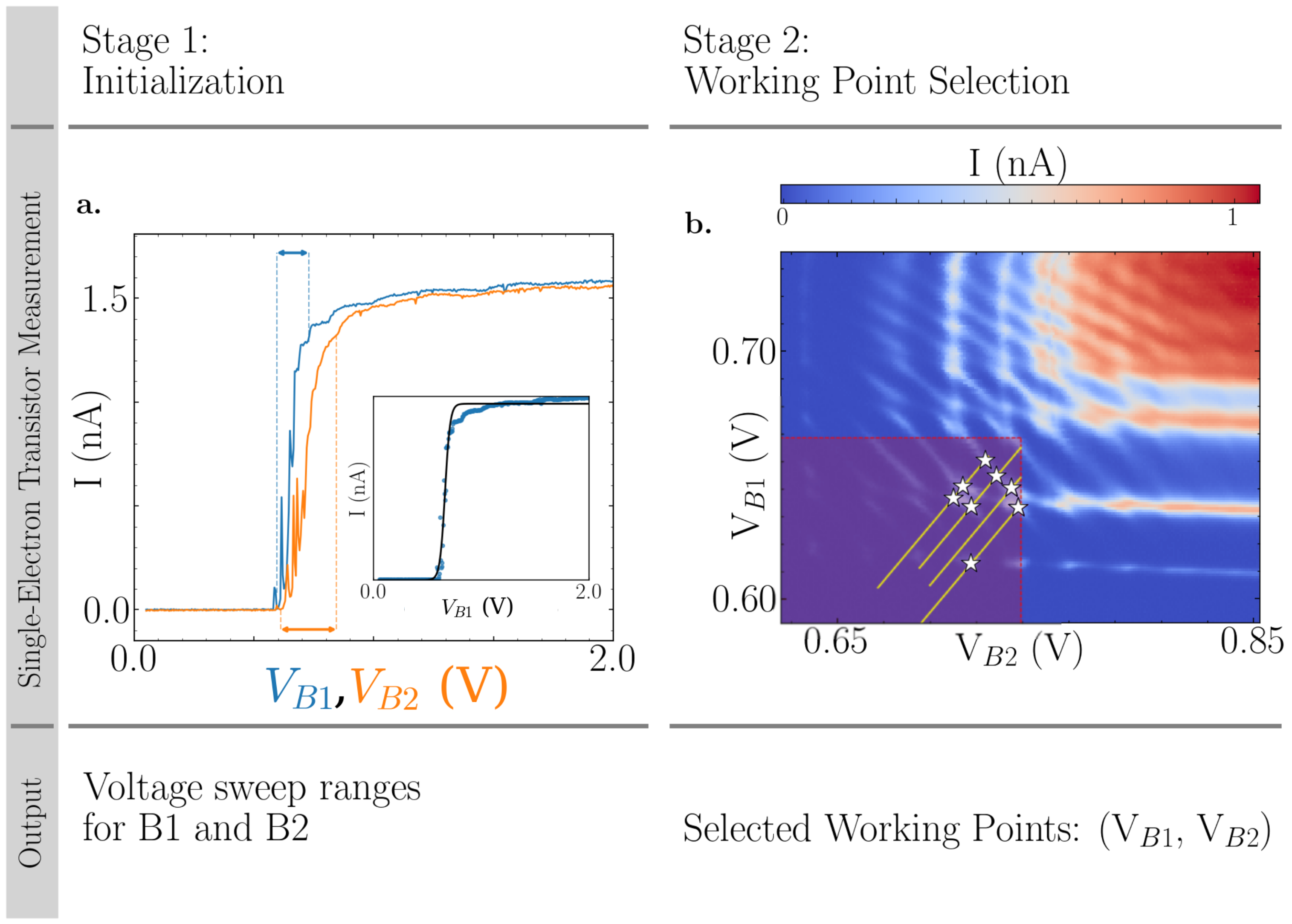}
    \caption{Overview of the autotuning protocol demonstrated on the SET at a temperature of 1.5~K. Stage 1 involves global turn-on followed by \textbf{(a)} lowering barrier gate voltages separately to determine  their pinch-off curves, which are fitted to a sigmoid function. Source-drain bias is 100 $\mu V$. The output of stage 1, indicated by the blue and orange arrows, is the center position and characteristic width of each sigmoid. The inset illustrates the sigmoidal fit to the B1 pinchoff curve. \textbf{(b)} In stage 2, a 2D current-voltage measurement is performed based on the barrier gate voltage ranges determined in stage 1. The pinch-off region is shown as shaded purple, and is the standard range used when running the protocol, although a 2D sweep over a wider range is shown here for clarity. To choose candidate working points $(V_{\mathrm{B1}},V_{\mathrm{B2}})$ in the sequential tunneling regime, image analysis tools are used to detect Coulomb oscillation features and define a set of 1D current traces along lines perpendicular to these features (indicated by the yellow parallel lines). Up to four current peaks along each of these 1D traces (indicated by stars) are selected as candidate working points.}
    \label{fig:overview}
\end{figure}

Briefly, the tuning protocol consists of three main stages: initialization, working point selection, and sensitivity tuning, as depicted in Figures~\ref{fig:overview} and~\ref{fig:Stage 3}. In the first stage, all gate voltages initialized at 0 V are swept to positive (negative) values simultaneously until global turn-on is established for the SET (SHT). The two barrier gates are then individually swept back down to verify that they can pinch-off the conducting channel and to characterize gating asymmetry. Following this stage, a two-dimensional scan of current versus the two barrier gates is obtained in the vicinity of the barrier pinch-off threshold. Coulomb oscillations are detected and operating voltage values for B1 and B2 are selected to form the ``working point'' $(V_{\mathrm{B1}},V_{\mathrm{B2}})$. In the final stage, the plunger gate P is swept while the barrier gates are fixed at the selected working point, yielding a one-dimensional SET (or SHT) current trace exhibiting Coulomb oscillations. Optimal charge sensing points $(V_{\mathrm{B1}},V_{\mathrm{B2}}, V_{\mathrm{P}})$, which we call ``operating points", are identified and ranked. The following subsections describe each stage in detail, using the SET device as a demonstrator.

\subsection{Initialization}
All gate electrodes are initially grounded, and a small source-drain voltage $V_{SD}$ (typically 100 $\mu$V) is applied to the ohmic contacts of the SET. This is followed by verifying that a current channel can form in the device by sweeping all gate voltages simultaneously positive (negative) until carriers accumulate and the device conducts (note, in the present devices the plunger side-gate P is kept grounded at this stage). The sweep is halted when a preset current threshold $I_{\text{th}}=1.5$ nA is met. If the current does not reach the threshold value, the sweep is halted at a preset maximum gate voltage and the device is rejected. The I-V curve from this sweep is fitted to the rectified linear unit, $\mathrm{ReLU}(V) = \max(0,\,a(V - V_{turnon}))$, where $V_{turnon}$ is the estimated turn-on voltage and a is the slope of the linear unit. These are used to define the parameters for the subsequent sweep. 

After global turn-on, the next step involves determining whether the barrier gates are able to pinch-off the channel current. Failure of a barrier gate to reach pinch-off could be due to imperfect fabrication or electrostatic discharge damage, and would warrant rejection of the device. While holding the other gates at the global saturation voltage, each barrier gate is independently swept down, initially to the turn-on voltage found in step 1. If pinch-off is not observed, the gate sweep range is extended downwards in increments of 0.25 V, iteratively, until pinch-off occurs or until a predetermined safe voltage differential is reached. Pinch-off is defined by the device current falling below a minimum threshold. This iterative approach prevents the voltage differences between gates from exceeding a safe value. 

An example of this measurement is shown in Figure~\ref{fig:overview}\textbf{(a)}, where the two barrier gate sweeps are overlaid. The outputs of the first stage, as indicated in Figure~\ref{fig:overview}, are the voltage ranges for each barrier around which pinch-off occurs. The data are fit to the logistic function $I(V) = I_0 + \frac{I_{max}-I_0}{1+e^{-b(V-V_0)}}$, where $I_{max}$ is the maximum current, $V_{0}$ is the inflection point of the sigmoidal curve, and the parameter $b$ is proportional to the slope at the inflection point. The region within $V_0 \pm \frac{\sqrt{8}}{b}$ is chosen as the voltage range of interest corresponding to the pinch-off width. This interval corresponds to the central transition region of the fitted sigmoid from $5\%$ to $95\%$ of the step amplitude. Limiting the two-dimensional sweep in stage 2 to being inside this range ensures that the device is operated in a safe regime where a quantum dot is expected to form, since sequential tunneling and confinement occur near pinch-off. Coulomb oscillations in the device current are seen near pinch-off for both barrier sweeps in Figure~\ref{fig:overview}\textbf{(a)}, likely due to spurious dots forming under or near the barrier gates. The pinch-off measurement for the SHT device at $\approx 50$ mK is shown in Figure~\ref{fig:SHT}\textbf{(a)}. Reasonable sigmoidal fits are still obtained despite the lower temperature and corresponding sharper Coulomb oscillation features. 

\subsection{Working point selection}
The second stage of the tuning protocol entails a 2D I-V measurement in which both barrier gates are swept within well-defined ranges near pinch-off, as determined by the output of stage 1. Each barrier gate voltage ranges from the lower limit $(V_0 - \frac{\sqrt{8}}{b})$ to the inflection point $(V_0)$ of the sigmoidal fit. The measurement shown in Figure~\ref{fig:overview}\textbf{(b)} was taken over a larger range $(V_0 \pm \frac{\sqrt{8}}{b})$ for clarity. Coulomb oscillations due to the intended SET dot are observed as diagonal lines appearing in the middle to lower left of the figure, oriented at a roughly -45 degree angle. The red dotted lines indicate the inflection point $(V_0)$ values for each barrier gate. The horizontal and vertical sharp features likely arise from disorder-induced regions of confinement below the barrier gates. The angle of the diagonal Coulomb oscillation lines from the intended dot relative to the horizontal depends on the ratio of capacitive coupling of the dot to the source and drain as well as the barrier voltage ranges. It can be inferred that the SET quantum dot in this case formed nearly symmetrically between the two barriers. 

\begin{figure}[t]
    \centering
    \includegraphics[width=\linewidth]{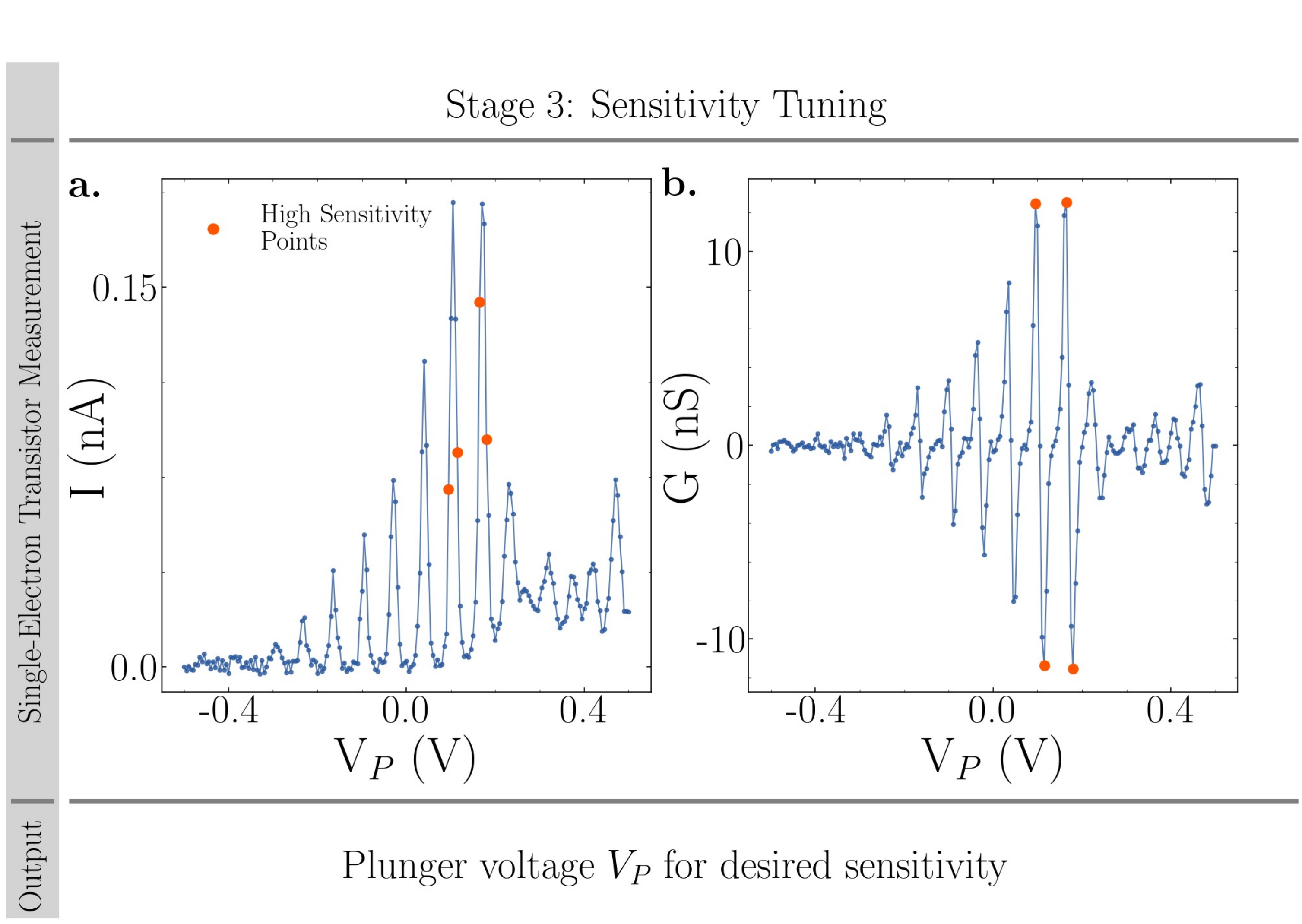}
    \caption{Stage 3 of the tuning protocol performed on the SET device at a temperature of 1.5~K. \textbf{(a)} A representative 1D current trace for the SET as a function of the plunger gate voltage, with a source-drain bias of 100 $\mu V$. Each such current trace is taken with the barrier gate voltages set to working points found in Stage 2. Within each trace, the highest sensitivity points are found by computing the corresponding transconductance G. \textbf{(b)} The transconductance corresponding to the 1D current trace in \textbf{(a)}. The 4 points with highest transconductance, up to the resolution of the measured trace, are labeled with orange dots.}
    \label{fig:Stage 3}
\end{figure}

For the tuning protocol to select a suitable working point in barrier-barrier space, the gradient of the current data is calculated, normalized, then passed through a band threshold to distinguish the Coulomb oscillation regime from both the noise floor background and areas of high current. Ridge detection is then used to identify Coulomb oscillation features ~\cite{Sato1998}. This is done by defining a ``ridgeness" value, stemming from the eigenvalues of the Hessian matrix at each pixel. These values are then used as inputs for line segments found by applying a probabilistic Hough transform~\cite{Galamhos1999}. The line segments are then filtered for suitable angles, for example, $-35^\circ \leq \theta \leq-55^\circ$. For each such line segment, a perpendicular line oriented at $\theta + 90^\circ$ is defined running through midpoint of the line segment (yellow lines in Figure~\ref{fig:overview}\textbf{(b)}). 1D current traces are taken from the original 2D current data along these perpendicular lines. Up to four of the highest current peaks along these traces are detected (star symbols in Figure~\ref{fig:overview}\textbf{(b)}).  

\subsection{Sensitivity tuning}
\begin{figure}[H]
    \centering
    \includegraphics[width=\linewidth]{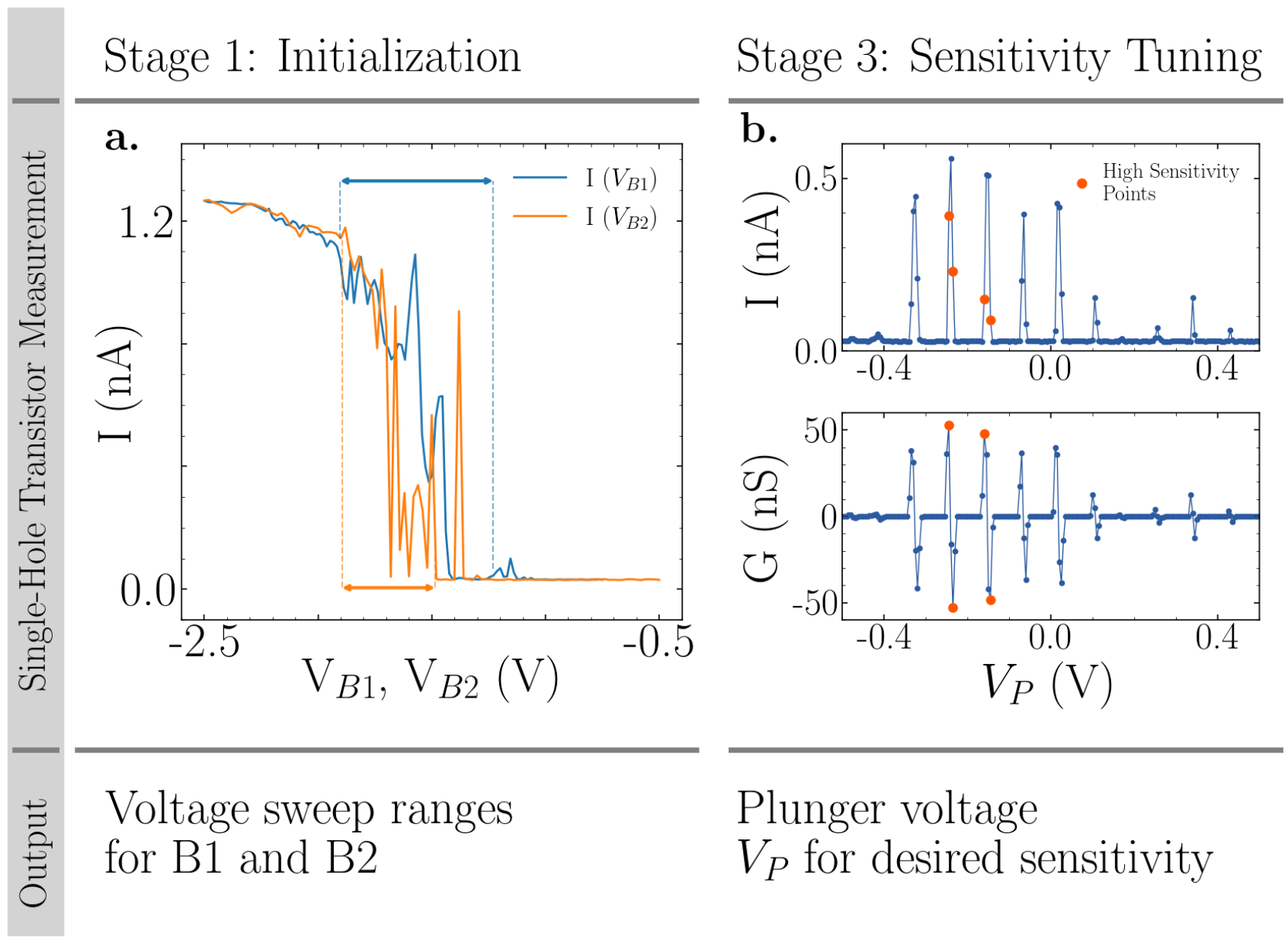}
    \caption{Stages 1 and 3 of the tuning protocol performed on the SHT device at $T \approx 50$ mK. \textbf{(a)} Pinch-off curves; the double-sided arrows indicate the widths of the sigmoidal fits in stage 1. \textbf{(b)} Charge sensor operating points found for the SHT in stage 3. The orange dots indicate the four highest sensitivity points in the 1D trace shown.}
    \label{fig:SHT}
\end{figure}

The third stage of the tuning protocol performs a sweep of the plunger side-gate voltage $V_{\mathrm{P}}$ at the working points selected in stage 2. This results in a set of 1D current traces exhibiting well-defined Coulomb oscillations if a quantum dot has successfully formed, as shown in Figure~\ref{fig:Stage 3}\textbf{(a)}. A predetermined range is first used for the plunger gate sweep, for example, $\pm 0.5$~V. This range can be adjusted based on a desired number of Coulomb oscillations using peak extraction. Its derivative is calculated numerically via the central difference, in order to obtain the transconductance $G (V_{\mathrm{P}}) = \mathrm{d}I/\mathrm{d}V_P$. Peaks in the transconductance are detected and sorted based on their magnitudes. The $V_{\mathrm{P}}$ values corresponding to the four highest sensitivity points (for each of the four working points) are stored and ranked. Examples are shown in Figure~\ref{fig:Stage 3}\textbf{(b)} for the SET and Figure~\ref{fig:SHT}\textbf{(b)} for the SHT. 

High sensitivity points are typically used for charge sensing related to qubit readout or noise spectroscopy. In the latter, gate voltages are held constant and the sensor current is recorded over a large number of time points. For time step $\Delta t$ and $N$ data points, a Fourier transform yields a noise power spectral density (PSD) over a frequency range $f \in [\frac{1}{N\Delta t}, \frac{1}{2\Delta t}]$~\cite{Freeman_2016, Petit_2018, Stuyck_2020}. PSDs taken at both high and low sensitivities can be compared in order to distinguish intrinsic noise originating from the electrostatic environment of the charge sensor versus extrinsic noise from the measurement setup or cable pickup. 

\subsection{Coulomb Diamond Analysis}

\begin{figure}[ht]
    \centering
    \includegraphics[width=1\linewidth]{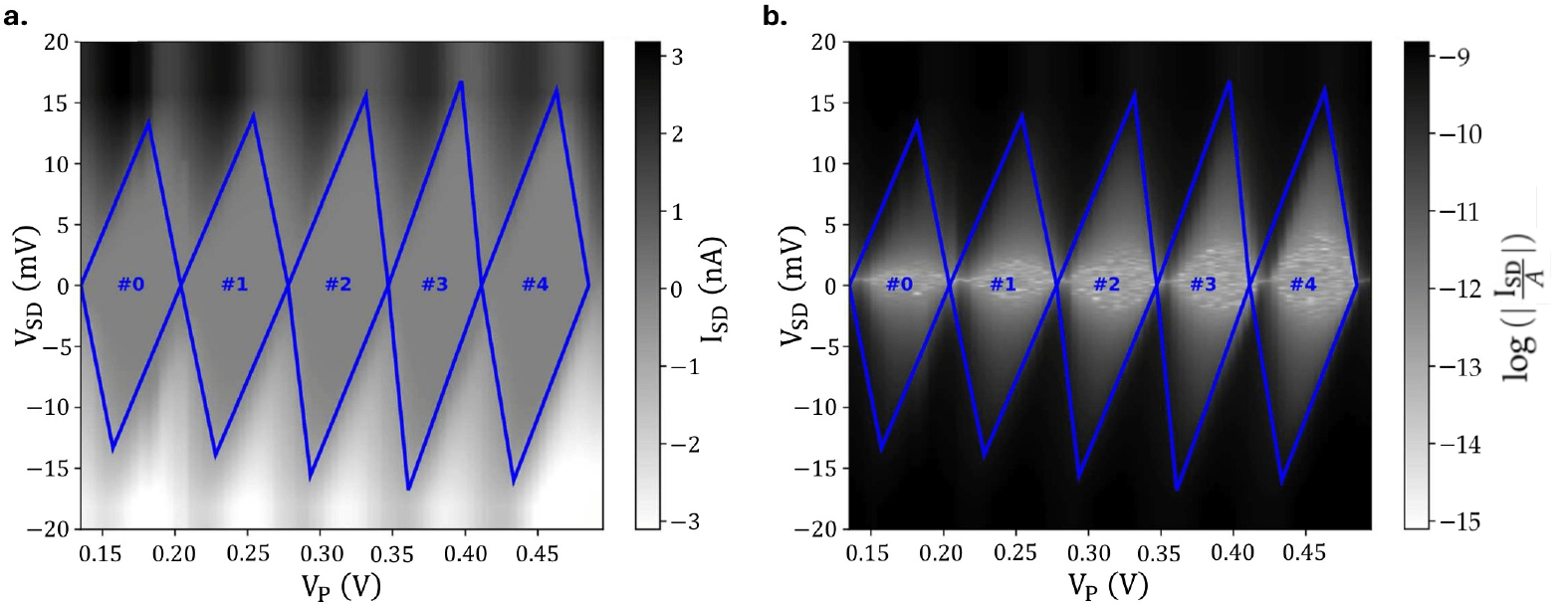}
    \caption{Coulomb diamonds for the SET measured at 1.5~K, with detected edges overlaid in solid blue lines. \textbf{(a)} Source-drain current $I_{SD}$ indicated in colorscale, versus the plunger gate voltage ($V_P$) and source-drain voltage ($V_{SD}$). \textbf{(b)} Detected diamond edges plotted over the logarithm of the current.} 
    \label{fig:diamonds}
\end{figure}

In addition to tuning, we have extended the protocol to characterization of SET/SHT properties. Key parameters can be determined by analyzing Coulomb diamonds, wherein varying both the plunger and source-drain voltages generates a 2D dataset, as depicted in Figure~\ref{fig:diamonds}~\textbf{(a)}. This sweep is performed at the working point $(V_{\mathrm{B1}},V_{\mathrm{B2}})$ determined in stage 2, and over the same range of $V_{\mathrm{P}}$ used for sensitivity tuning. To establish the sweep range for the source-drain voltage, a value is estimated based on the charging energy of a circular dot with a diameter equal to the spacing between the two barrier gates. A buffer of 25~\% is added to account for variations in dot size.

To automate the detection and analysis of Coulomb diamonds, the data is first filtered by taking the logarithm of the current and applying a mask based on a preset binary threshold. This results in an image where each pixel corresponds to one of two distinct values, separating the low conductance diamond regions from the high conductance regions outside the diamonds. A Gaussian blur is then applied to the image in order to smooth the edges and reduce noise. The image is segmented by the horizontal line corresponding to zero source-drain bias, and a Hough transform is performed on each of the two halves to detect the lines corresponding to diamond edges. For each diamond, we average the slopes for each pair of edges with the same sign of slope. From the intersections of opposite-sign edges we extract the diamond vertices, from which we obtain the diamond width in $V_P$ and half-height in $V_{SD}$ used to calculate $\alpha$ and the capacitances. Figure~\ref{fig:diamonds}~\textbf{(b)} shows the detected diamond edges overlaid on $\log(|I_{SD}|)$, highlighting the clarity in distinguishing the diamond shapes when using the logarithm of the current. A slight misalignment in the intersection points of the detected edges with respect to the underlying data may be due to the low resolution of the image; a voltage sweep with finer spacing should yield higher resolution and more accurate line detection by the Hough transform. Accuracy is also dependent on the choice of binary threshold; in practice we choose a threshold that robustly separates the blockade regions from the conducting regions.

These fits to the diamonds are next used to calculate key device parameters, for both the SET and SHT devices used in this study. The results are summarized in Table~\ref{tab:parameter_table}. The gate voltage lever arm $\alpha$ is given by the ratio of the diamond half-height in bias, converted to energy via $e$, to the diamond width in gate voltage. The Coulomb period, charging energy, and dot capacitances are determined via the Constant Interaction model, where the total capacitance of a single dot is the sum of the gate, source, and drain capacitances \cite{leo_dot,Hanson_2007}. The lateral extent of the dot is estimated by modeling the dot as a circular parallel plate capacitor, from which the radius $r$ can be calculated \cite{radius}. These parameters are useful for characterizing the SET/SHT as a charge sensor and verifying that such a quantum dot reasonably matches expectations in terms of size and capacitance. For example, it can be seen from this analysis that the estimated dot radii are in the range of $30 - 36$ nm, which matches expectations for the barrier gate separation of $60$ nm. Given that the SET and SHT are very similar in structure and were measured under different experimental conditions, it is encouraging that the properties summarized in Table~\ref{tab:parameter_table} are comparable.

\begin{table}[h!]
    \centering
    \begin{tabular}{|c|c|c|c|c|}
    \hline
    \textbf{Parameter} & \textbf{Symbol} & \textbf{SET Values} & \textbf{SHT Values} & \textbf{Unit} \\
    \hline
    Lever Arm & $\alpha$ & $211 \pm 4$ & $141 \pm 6$ & meV/V \\
    \hline
    Coulomb Period & $\Delta V_P$ & $70.0 \pm 0.7$ & $81.1 \pm 1.6$ & mV \\
    Charging Energy & $E_C$ & $14.7 \pm 0.2$ & $11.3 \pm 0.4$ & meV \\
    \hline
    Total Capacitance & $C_\Sigma$ & $11.0 \pm 0.2$ & $14.4 \pm 0.5$ & aF \\
    Gate Capacitance & $C_G$ & $2.29 \pm 0.02$ & $1.99 \pm 0.04$ & aF \\
    Source Capacitance & $C_S$ & $3.0 \pm 0.1$ & $4.9 \pm 0.2$ & aF \\ 
    Drain Capacitance & $C_D$ & $5.40 \pm 0.08$ & $7.50 \pm 0.2$ & aF \\
    \hline
    Dot Radius & $r$ & $29.8 \pm 0.3$ & $36.3 \pm 0.6$ & nm \\
    \hline
    \end{tabular}
    \caption{Parameters of the measured SET and SHT devices estimated based on the automated collection and analysis of Coulomb diamonds. Values listed are averages across each set of detected diamonds, and errors are the standard deviations. Note that $E_C = \alpha \Delta V_P$, where $\Delta V_P$ is the Coulomb period measured in plunger gate voltage.}
    \label{tab:parameter_table}
\end{table}

\section{Discussion}

This work demonstrates a practical end-to-end procedure for bringing up and operating MOS single-electron and single-hole transistors as DC charge sensors with minimal device-specific input. The protocol (i) initializes an unknown device after cooldown, (ii) identifies barrier-gate working points that lie in a Coulomb blockade regime appropriate for sensing, (iii) selects and ranks high-sensitivity operating points using a fast proxy metric (transconductance), and (iv) optionally performs automated Coulomb-diamond acquisition and analysis to extract physically meaningful device parameters (lever arms, capacitances, and estimated dot size). Importantly, we demonstrate the same workflow on both an accumulation-mode SET at 1.5~K and an analogous SHT at dilution refrigerator temperatures (Figures~\ref{fig:overview} -~\ref{fig:SHT}), indicating that the same protocol can be applied to both device polarities and across a wide temperature range.

\subsection{Implications for charge sensing at elevated temperature}

A central result is that the SET can be tuned to a high-sensitivity charge-sensing regime at 1.5~K (Fig.~\ref{fig:overview} and  Fig.~\ref{fig:Stage 3}), provided that the dot energy scales remain large compared to thermal broadening. For compact MOS dots, the charging energies inferred from Coulomb diamonds (Table~\ref{tab:parameter_table}) are on the order of \(\sim 10\)--\(15\)~meV, far exceeding \(k_{\mathrm{B}}T\) at 1--2~K (\(\sim 0.1\)--\(0.2\)~meV). This margin helps explain why well-defined Coulomb oscillations and steep transconductance features persist at 1.5~K, and it supports the feasibility of DC charge sensing (and DC pre-tuning for RF-SET operation) in the 1--2~K regime for sufficiently compact devices. Even when $E_C \gg k_B T$, the achievable transconductance for sensing can also be limited by lifetime (tunnel) broadening set by the dot--lead tunnel rates; in our protocol we mitigate this by selecting barrier-gate working points near pinch-off in the sequential-tunneling regime, avoiding overly transparent barriers. From a systems perspective, extending reliable charge sensing upward in temperature is relevant to scalable architectures in which refrigeration power, wiring density, and proximity to cryogenic control electronics motivate operation above 1 Kelvin.

\subsection{Protocol performance, bottlenecks, and robustness}

A key practical benefit of automation is reduction of hands-on tuning time and improved repeatability across cooldowns and devices. In the present implementation, the dominant runtime cost is the stage~2 barrier--barrier 2D sweep (Fig.~\ref{fig:overview}~\textbf{(b)}), which takes \(\sim 25\)~minutes for the example shown, a plot of 150 x 230 points, with a settling time of 25 ms, while the remaining stages together typically require \(\lesssim 5\)~minutes. This bottleneck is largely set by the control/readout stack (DC sources, DMM acquisition, and GPIB communication). Modern digitizers and FPGA-based control can substantially reduce this overhead by (i) acquiring current traces at much higher sample rates and transferring buffered blocks to the host computer and (ii) generating fast, repeatable voltage ramps on multiple gates, with the remaining bandwidth limitation often set by the low-noise current preamplifier \cite{FPGA_acc}. Even without changing the device physics, hardware modernization alone could reduce total protocol time by orders of magnitude.

Beyond hardware speed, measurement time can be reduced by sampling less densely while retaining the information needed for working-point selection. The 2D barrier map offers a global view of the stability landscape, revealing the intended Coulomb-ridge family while exposing competing features such as disorder-induced dots, lead structure, or multi-dot formation. In many cases, similar information can be obtained from a small set of targeted 1D sweeps along selected directions in \((V_{\mathrm{B1}},V_{\mathrm{B2}})\) space (e.g., approximately perpendicular to the ridges, with a few nearby angles added as needed), using an adaptive scheme that expands the set only when the initial cuts are ambiguous. A related reduction is plausible for Coulomb diamonds: instead of a dense 2D sweep in \((V_P,V_{SD})\), a limited number of \(V_P\) traces at selected \(V_{SD}\) values may suffice to reconstruct diamond edges by tracking peak trajectories.

Robustness of working-point selection is governed by both device disorder and algorithmic assumptions. In the present protocol, ridge detection and Hough transforms identify candidate Coulomb features and then restrict accepted line segments to a window of angles (e.g., \(-35^\circ \le \theta \le -55^\circ\) in Fig.~\ref{fig:overview}~\textbf{(b)}). This angle filter is motivated by the expected geometry of a roughly symmetric double-barrier dot, where the intended ridge family appears approximately diagonal near pinch-off, and it helps suppress horizontal/vertical features commonly associated with barrier-localized confinement or abrupt channel turn-on. However, ridge angles can vary across devices and operating conditions, and multi-dot regimes can produce multiple slope families. A more general implementation could infer the dominant ridge angle(s) directly from the map and adapt the acceptance window accordingly, reducing reliance on fixed device-specific expectations while preserving the speed of the current approach.

Certain pathological regions of the barrier map warrant explicit treatment because they can mislead working-point selection if one simply seeks large current. For example, broad anti-crossing-like features or high-conductance patches can occur in regimes where an unintended dot forms or where transport is dominated by mechanisms not favorable for stable sensing. In the present protocol, candidate selection already penalizes points in very high-current regions and discourages choices near such broad features. Additional robustness could come from incorporating morphology-based scores (e.g., preference for thin, ridge-like features of consistent slope and suppression of wide bright regions) rather than relying primarily on local conductance.

\subsection{Sensitivity metric and operating-point ranking}

In stage~3, operating points are ranked by the magnitude of the transconductance \(|\mathrm{d}I/\mathrm{d}V_P|\), which is a fast and convenient proxy for charge sensitivity because it uses the same 1D trace already acquired (Fig.~\ref{fig:Stage 3}~\textbf{(b)} and Fig.~\ref{fig:SHT}~\textbf{(b)}). In many experiments, however, the relevant figure of merit is the signal-to-noise ratio within a given bandwidth, which depends both on slope and on the current noise \(S_I\). A natural extension of the protocol is therefore to augment the ranking with a short dwell measurement at each candidate peak to estimate the noise floor and compute an effective score proportional to \(|\mathrm{d}I/\mathrm{d}V_P|/\sqrt{S_I}\) over a fixed bandwidth. This modification would better reflect the practical performance of the sensor for time-domain readout and for noise spectroscopy, while adding only modest measurement overhead compared to a full 2D scan.

\subsection{Automated Coulomb-diamond analysis as screening and calibration}

Automated Coulomb-diamond acquisition and fitting (Fig.~\ref{fig:diamonds}) provides a second layer of value beyond simply finding a sensing point. First, it enables rapid device screening: lever arms and charging energies quantify whether a sensor is likely to operate at elevated temperature and whether its energy scales match expectations for the device geometry. Second, the extracted capacitances and inferred dot size (Table~\ref{tab:parameter_table}) serve as diagnostics of dot placement and barrier symmetry; for example, asymmetries in \(C_S\) and \(C_D\) can indicate a dot displaced toward one lead, which may correlate with ridge-angle shifts in barrier maps and with susceptibility to disorder features. We note that the extracted dot radius should be interpreted as an effective electrostatic size under a simplified model, useful primarily for screening and device-to-device comparison. Third, lever arms provide the calibration needed to convert voltage noise to energy noise, which is useful when comparing devices or when interpreting noise spectroscopy data.

There are also clear limitations to the present automated diamond extraction. The line-detection accuracy depends on the resolution of the \((V_P,V_{SD})\) sweep and on the binary threshold used to segment low- and high-conductance regions; insufficient resolution or an ill-chosen threshold can shift detected intersections and bias slopes. Moreover, the Constant Interaction model provides a convenient and widely used framework for parameter extraction, but real devices can show diamond distortions from inelastic tunneling processes or lead density-of-states structure. Future improvements could incorporate (i) resolution-aware uncertainty estimates on fitted slopes, (ii) automated threshold selection (e.g., by histogram-based methods) or edge detection on \(\mathrm{d}I/\mathrm{d}V_P\) rather than \(\log|I|\), and (iii) model-based fitting that explicitly allows for asymmetries in tunnel coupling and capacitances. Even in its current form, however, a consistent automated analysis is valuable when comparing large device sets where manual fitting does not scale.

\subsection{Integration into a larger automation and readout stack}

The protocol can serve as a standardized cooldown-to-sensor bring-up step: it verifies channel formation and pinch-off, selects a charge-sensing operating point, and extracts lever arms and energy scales for subsequent tuning. These outputs provide consistent starting conditions for downstream routines such as qubit-dot formation, virtual gating, and readout optimization. When longer unattended runs are required, the operating point can be maintained using established feedback schemes \cite{dynamic_feedback}, and stability-diagram classifiers trained on large datasets (e.g., \emph{QFlow} \cite{qflow}) can be used to flag multi-dot or disorder-dominated regimes \cite{AndrijaThesis}.

\section{Conclusions and Outlook}

We have presented an automated protocol for tuning silicon MOS single-electron and single-hole transistors to operate as sensitive DC charge sensors. Starting from a post-cooldown state with minimal device-specific information, the protocol reliably initializes the device, identifies a barrier-gate working point in the Coulomb blockade regime, and selects and ranks high-sensitivity operating points based on transconductance. In parallel, the same framework enables automated acquisition and analysis of Coulomb diamonds to extract sensor-relevant parameters, including gate lever arms, charging energies, capacitances, and an estimated dot size.

We demonstrated the approach on accumulation-mode SET and SHT devices measured at 1.5~K and \(<100\)~mK, respectively, showing that the method is ambipolar and applicable across a wide temperature range. The successful tuning and characterization at 1.5~K highlights the feasibility of charge sensing in the 1--2~K regime for sufficiently compact MOS devices, supporting efforts toward higher-temperature operation relevant to scalable spin-qubit platforms.

Future work will focus on reducing runtime and improving robustness. On the measurement side, faster control and acquisition hardware and/or adaptive sampling strategies can substantially shorten the time-intensive 2D sweeps. Reliability of analysis can be improved by incorporating noise-aware operating-point ranking, machine-learning-based identification of desirable single-dot sensor regimes, and dynamic feedback to maintain selected operating points in the presence of slow charge drifts. Together, these developments would further position automated charge-sensor bring-up and calibration as reusable infrastructure that can integrate directly with broader quantum-dot and qubit tuning protocols.

\section*{Acknowledgments}

This research was undertaken thanks in part to funding from the Canada First Research Excellence Fund (Transformative Quantum Technologies) and the Natural Sciences and Engineering Research Council (NSERC) of Canada. This infrastructure would not be possible without the significant contributions of CFREF-TQT, CFI, ISED, the Ontario Ministry of Research and Innovation, and Mike and Ophelia Lazaridis. The authors acknowledge support from the Netherlands Organization of Scientific Research (NWO) under VICI grant VI.C.222.083, European Union's Horizon 2020 and Horizon Europe Research and Innovation Programme under Grant Agreement No. 951852 (QLSI) and No. 101080022 (ONCHIPS).

\section*{Author contributions}
A.P. developed the tuning protocol and performed the experiments with T.J. and B. V. O. A.S., T.J., B. V. O. and J.B. wrote the manuscript. D.B. and Q.T.N. fabricated the devices. F.A.Z. and J.B. contributed equally to supervising the project. 

\section*{Data availability statement}
The data that support the findings of this study are available upon request from the authors. 

\section*{ORCID iDs}
\noindent Jonathan Baugh https://orcid.org/0000-0002-9300-7134

\section*{References}

\bibliography{references}



\end{document}